\documentclass[reprint,aps,prd,floatfix,nofootinbib,showkeys,showpacs]{revtex4-2}
\usepackage{amsmath,amsthm,amssymb}
\usepackage{graphicx,color,dsfont,bm, url, hyperref}
\usepackage[utf8]{inputenc}
\def\C{\mathcal{C}}
\def\F{\mathcal{F}}
\def\W{\mathcal{W}}

\def\inn{\mathrm{in}}
\def\out{\mathrm{out}}
\def\X{{\scriptscriptstyle\mathrm{X}}}
\def\R{{\scriptscriptstyle\mathrm{R}}}
\def\V{{\scriptscriptstyle\mathrm{V}}}

\def\ktt{\scriptscriptstyle\mathrm{k_t}}

\def\CF{\mathrm{C_F}}
\def\CFsq{\mathrm{C_F^2}}
\def\CFcub{\mathrm{C_F^3}}
\def\CFfour{\mathrm{C_F^4}}
\def\CFfive{\mathrm{C_F^5}}
\def\CFsix{\mathrm{C_F^6}}

\def\s{\sigma}
\def\S{\Sigma}
\def\oS{\overline{\Sigma}}
\def\O{\Omega}
\def\Ob{\bar{\Omega}}

\def\as{\alpha_s}
\def\asb{\bar{\alpha}_s}
\renewcommand{\d}{\mathrm{d}}
\def\qb{\bar{q}}
\def\uh{\hat{u}}
\def\Uh{\hat{\mathcal{U}}}
\def\kt{$k_t$ }

\def\oXi{\overline{\Xi}}

\begin{document}

\title{Clustering logarithms up to six loops}

\author{K. \surname{Khelifa-Kerfa}}
\affiliation{Department of Physics, Faculty of Science and Technology\\
Relizane University, Relizane 48000, Algeria}
\email{kamel.khelifakerfa@univ-relizane.dz}

\begin{abstract}
We compute the leading clustering (abelian non-global) logarithms, which arise in the distribution of non-global QCD observables when final-state partons are clustered using the \kt jet algorithm, up to six loops in perturbation theory. Our calculations are based on the recently introduced formula for the analytic structure of \kt clustering \cite{Khelifa-Kerfa:2024roc}. These logarithms exhibit a pattern of exponentiation and are subsequently resummed into an exponential form. We compare this resummed result with all-orders numerical calculations.
\end{abstract}


\maketitle

\section{Introduction}
\label{sec:Intro}

The intricate dynamics of QCD are vividly manifested in the formation and evolution of jets at high-energy colliders, such as the Large Hadron Collider (LHC). Jets--collimated sprays of hadrons arising from the fragmentation of final-state partons--are indispensable tools for probing the structure of QCD across both perturbative and non-perturbative scales. Theoretical calculations of jet observables, such as jet shapes and substructure, demand a nuanced understanding of QCD radiation patterns, governed by the interplay of soft and collinear emissions.

Non-global observables constitute a class of QCD observables that depend on the energy flow within restricted regions of phase space. These observables pose formidable theoretical challenges. Unlike their global counterparts, their distributions are shaped by the intricate interplay of soft radiation both within and outside the measured region. This interplay gives rise to large logarithms, known as non-global logarithms (NGLs) \cite{Dasgupta:2001sh}, which disrupt the perturbative series. NGLs encapsulate correlations between soft, wide-angle gluons emitted across disparate energy scales and angular regions, making their precise calculation critical for advancing the understanding of jet shapes and substructure. Over recent years, NGLs have been extensively studied (see, for instance, \cite{Banfi:2002hw, Banfi:2010pa, Khelifa-Kerfa:2011quw, Dasgupta:2012hg, Delenda:2012mm, Hatta:2013iba, Khelifa-Kerfa:2015mma, Schwartz:2014wha, Banfi:2021owj, Becher:2023vrh}).

Another subclass of non-global logarithms, referred to as clustering logarithms (CLs), arises intrinsically from the application of jet algorithms, which are fundamental to experimental jet reconstruction. These logarithms, associated with large contributions from the clustering of abelian, independent primary gluon emissions, are often termed abelian non-global logarithms. CLs were first identified in \cite{Banfi:2005gj} in the context of the \kt jet algorithm \cite{Catani:1993hr, Ellis:1993tq}, where it was demonstrated that while clustering suppresses the impact of standard NGLs, as earlier established in \cite{Appleby:2002ke}, it simultaneously generates a hierarchy of CLs at every perturbative order in the non-global observable’s distribution. Consequently, an accurate determination of CLs is indispensable for precision computations.
The all-orders resummation of leading CLs has, until recently, been achieved solely through the Monte Carlo program of \cite{Dasgupta:2001sh}, implemented specifically for the \kt algorithm. However, recent advancements reported in \cite{Becher:2023znt} have numerically resummed leading CLs within the framework of Soft and Collinear Effective Theory (SCET) for the gap-between-jets observable, encompassing various sequential recombination jet algorithms, including $k_t$.

Fixed-order calculations of CLs at the leading logarithmic level were conducted in the original work \cite{Banfi:2005gj} for the gap-between-jets (interjet-energy-flow) observable, employing the \kt clustering sequence, at two loop order. Similar calculations were later computed for various other observables (see, for instance, \cite{Banfi:2010pa, Khelifa-Kerfa:2011quw, Kerfa:2012yae, Ziani:2021dxr, Bouaziz:2022tik, Benslama:2023gys}). These calculations were subsequently extended to four loops in \cite{Delenda:2006nf}. Ref. \cite{Delenda:2012mm} presented analogous computations for the single-jet mass observable, incorporating results for both the $k_t$ and Cambridge/Aachen jet algorithms \cite{Dokshitzer:1997in, Wobisch:1998wt} up to four  and three loops, respectively. Furthermore, CLs coefficients beyond the leading-log were explicitly determined at two loops in \cite{Kerfa:2012yae}. All of these computations were carried out within the framework of eikonal (soft) theory, under the assumption of strong energy-ordering of the emitted gluons, which significantly simplified the calculations. However, the cited works adopted a brute-force approach to calculate CLs for these specific observables. Notably, no general methodology was available for systematically computing CLs for arbitrary observables at any loop order.

Recently, a generalized formula for the analytic structure of \kt clustering has been derived in \cite{Khelifa-Kerfa:2024roc}, featuring the following key attributes: (i) it enables simultaneous computation of both NGLs and CLs, (ii) it is applicable to any generic non-global observable, (iii) it allows calculations to be performed at any loop order (under the eikonal and strong energy-ordering approximations), (iv) it includes finite $N_c$ corrections, (v) it captures the complete jet-radius dependence, and (vi) it is valid across various high-energy processes, including both leptonic and hadronic collisions. This formula was recently employed in \cite{Khelifa-Kerfa:2024hwx} to calculate the full perturbative distribution of the dijet mass observable in $e^+ e^-$ annihilation processes at next-to-leading logarithm accuracy, up to four loops, incorporating both NGLs and CLs contributions.

In this paper, we extend the application of the formula from \cite{Khelifa-Kerfa:2024roc} to compute CLs at five- and six loop orders for a generic non-global observable. As shown in the main text, the clustering function (the component of the observable cross-section accounting for the effect of the \kt jet algorithm) factorizes at each loop order into two distinct terms: the first is a {\it reducible} component related to lower-order clustering functions, while the second is an {\it irreducible} term, which is genuinely new and not related to lower loop orders. This factorization pattern is subsequently reflected in the structure of the integrated cross-section.

To illustrate, we explicitly evaluate the integrals for the single-jet mass observable previously studied in \cite{Delenda:2012mm}. Our findings indicate that the CLs coefficients decrease significantly with increasing loop order, demonstrating rapid convergence of the perturbative series. Comparisons with Monte Carlo results from \cite{Dasgupta:2001sh} show good overall agreement. However, the five- and six loop contributions provide only a negligible improvement over the four loop result derived in \cite{Delenda:2012mm}, owing to the vanishingly small size of the corresponding CLs coefficients.

This paper is organized as follows. In Sec. \ref{sec:4loops}, we revisit the calculation of CLs for the single-jet mass observable up to four loops, reexamined in light of the master formula from \cite{Khelifa-Kerfa:2024roc}. Additionally, the necessary ingredients for the subsequent calculations are outlined. Sections \ref{sec:5loops} and \ref{sec:6loops} detail the new five  and six loop contributions to the CLs. In Sec. \ref{sec:AllOrders}, we compare these results with the Monte Carlo predictions of \cite{Dasgupta:2001sh} to evaluate the convergence of the fixed-order perturbative series up to six loops and to assess the impact of the missing higher-order terms. Finally, we present our conclusions in Sec. \ref{sec:Conclusion}.

\section{CLs up to four loops: recap}
\label{sec:4loops}
\subsection{Definitions}
\label{subsec:Definitions}

We consider the lepton process of $e^+ e^-$ annihilation into two final-state jets produced back-to-back in the threshold limit: $e^+ e^- \to j_q + j_{\bar{q}} + X$. Here, we assume that only the invariant mass squared of the quark-initiated jet, $j_q$, is measured, while the anti-quark-initiated jet, $j_{\bar{q}}$, is left unmeasured. Soft activity, $X$, outside the two jets may be restricted by an energy veto. Although the latter introduces large logarithms that can be computed straightforwardly (see, for instance, \cite{Banfi:2010pa, Khelifa-Kerfa:2011quw, Kerfa:2012yae}), these are not addressed in this work. At the partonic level, the process can be schematically expressed as follows:
\begin{align}\label{eq:Def:Process}
 e^+ + e^- \to q(p_a) + \qb (p_b) + g_1(k_1) + \cdots + g_n(k_n),
\end{align}
where $p_a$ and $p_b$ are the four momenta of the quark and anti-quark, respectively, and $k_i$ is the four-momentum of $i^{\text{th}}$ gluon. We work within the eikonal (soft) approximation, assume all partons to be massless ($p_i^2 = k_j^2 = 0$) and impose a strong energy-ordering condition on the emitted soft gluons: $Q \gg \omega_1 \gg \cdots \gg \omega_n$, where $Q$ represents the hard scale of the collision~\footnote{Alternatively, this ordering can be expressed in terms of the transverse momenta of the emitted gluons, using the relation $k_t = \omega \sin \theta$.}. Under these conditions, we achieve single logarithmic (SL) accuracy for the distribution of the invariant jet mass observable. At this accuracy, recoil effects can safely be neglected (see, for instance, \cite{Banfi:2004yd}).
The four-momenta described above can be parameterized as:
\begin{align}\label{eq:Def:FourMomenta}
 p_a &= \frac{Q}{2} \left(1,1 ,0,0\right), \qquad
 p_b = \frac{Q}{2} \left(1,-1,0,0\right), \notag\\
 k_i &= k_{ti} \left(\cosh \eta_i,\cos\phi_i,\sin\phi_i,\sinh\eta_i \right),
\end{align}
where $k_{ti}$, $\eta_i$ and $\phi_i$ are the transverse momentum, rapidity and azimuthal angles, respectively, of the $i^{\text{th}}$ gluon with respect to the beam axis.

We define the normalized invariant jet mass squared, $\varrho$, as:
\begin{align}\label{eq:Def:JetmassDef}
 \varrho &= \frac{4}{Q^2} \left( p_a + \sum_{k \in j_q} k_i \right)^2= \sum_{i \in j_q} \varrho_i + \mathcal{O} \left( \frac{k_{t}^2}{Q^2} \right) ,
 \notag\\
 \varrho_i &= 8 \frac{p_a \cdot k_i}{Q^2} = 2 x_i \left(\cosh \eta_i - \cos \phi_i\right),
\end{align}
where we have introduced the energy fraction $x_i = 2 k_{ti}/Q$ and neglected terms proportional to $k_t^2/Q^2$ in the soft limit. The jets are defined using the sequential recombination $k_t$ jet algorithm, as formulated in Ref.~\cite{Cacciari:2011ma}. We aim to compute the integrated jet cross-section (or equivalently the jet mass fraction) in the $k_t$ clustering, $\S^{\ktt}(\rho)$, defined as:
\begin{align}\label{eq:Def:JetmassXsec}
 \S^{\ktt}(\rho) &= \int \frac{1}{\s_0} \frac{\d \s}{\d \varrho} \Theta\left[\rho - \varrho(k_1, \dots, k_n) \right] \d \varrho,
\end{align}
where $\sigma_0$ is the Born cross-section, and $\frac{\d\sigma}{\d\varrho}$ is the differential cross-section of the process. Note that the mass (squared) of the measured quark jet, $\varrho(k_1, \dots, k_n)$, is restricted to be less than a value $\rho$, which we refer to as the jet veto. The presence of widely separated energy scales in this scattering process (such as $Q$ and $\rho$) results in the emergence of large logarithms in the ratio of these scales, as we shall later demonstrate. The fixed-order perturbative expansion of the jet mass fraction, Eq.~\eqref{eq:Def:JetmassXsec}, is given by:
\begin{align}\label{eq:Def:JetmassXsec-Series}
 \S^{\ktt}(\rho) = 1 + \S_1^{\ktt}(\rho) + \S_2^{\ktt}(\rho) + \cdots,
\end{align}
where the $m^{\text{th}}$ contribution in the above expansion is given by \cite{Khelifa-Kerfa:2015mma, Khelifa-Kerfa:2024roc}:
\begin{align}\label{eq:Def:JetmassXsec-m}
 \S_m^{\ktt}(\rho) &= \int_{x_1> \cdots >x_m} \left( \prod_{i=1}^m \d\Phi_i \right) \sum_\X \, \Uh_m \W_{12 \dots m}^\X,
\end{align}
where the sum is over all gluon configurations $X$, $\W_{12 \dots m}^X$ represents the squared eikonal amplitude for a particular configuration $X$, and $\Uh_m$ denotes the measurement operator at the $m$th order in the perturbation expansion. In the strong-energy ordering approximation, it factorizes into a product of single-soft-gluon measurement operators $\uh$ \cite{Schwartz:2014wha, Khelifa-Kerfa:2015mma, Khelifa-Kerfa:2024roc}:
\begin{align}\label{eq:Def:MeasOperator}
 \Uh_m = \prod_{i=1}^m \uh_i.
\end{align}
Detailed definitions and derivations of these terms are provided in our previous works (see, for instance, Refs. \cite{Khelifa-Kerfa:2015mma, Khelifa-Kerfa:2024roc}). In particular, the measurement operator for the $i$th soft emission may be recast as follows~\cite{Khelifa-Kerfa:2015mma, Khelifa-Kerfa:2024roc}:
\begin{align}\label{eq:def:uOperator}
 \uh_i = \Theta_i^\V + \Theta_i^\R \left[ \Theta_i^\out + \Theta_i^\inn \Theta\left( \rho - \varrho_i\right)  \right] = 1 - \Theta_i^\rho \Theta_i^\inn \Theta_i^\R,
\end{align}
where $\Theta_i^\V = 1$ if gluon $i$ is virtual and zero otherwise, $\Theta_i^\R = 1$ if gluon $i$ is real and zero otherwise, $\Theta_i^{\inn(\out)} = 1$ if gluon $i$ is in (out of) the jet region and zero otherwise, and $\Theta_i^\rho = \Theta(\varrho_i - \rho)$.
Therefore, applying $\uh_i$ to the eikonal amplitude squared, $\W_{12\dots m}^{\X}$, results in a non-vanishing contribution only if gluon $i$ is real, emitted inside the jet region, and satisfies the condition $\rho > \varrho_i$.
Notice that the last equality in Eq.~\eqref{eq:def:uOperator} follows directly from the identities:
$\Theta_i^\V+\Theta_i^\R = 1, \Theta_i^\inn + \Theta_i^\out = 1,$ and $\Theta_i^\rho = 1- \Theta(\rho - \varrho_i)$.

The phase space factor for the emission of gluon $k_i$ is expressed as:
\begin{align}\label{eq:Def:PhaseSpaceFactor}
 \d \Phi_i = \asb \, \frac{\d x_i}{x_i} \d\eta_i \frac{\d\phi_i}{2 \pi},
\end{align}
where $\asb = \as/\pi$. For the emission of $m$ energy-ordered {\it primary} gluons, the eikonal amplitude squared, $\W_{12 \dots m}^\X$, including the jet veto constraint, may be cast as:
\begin{align}\label{eq:Def:WP-m}
 \W_{1\dots m}^\X = -\left(-1\right)^{n_\V}\,\prod_{i=1}^{m} \CF\, w_{ab}^i,
\end{align}
with $n_\V$ being the number of virtual gluons in configuration X and the one loop eikonal antenna function, $w_{ab}^i$, is given by:
\begin{align}\label{eq:Def:1loopAntennaFunc}
  w_{ab}^i = k_{ti}^2 \frac{ \left(p_a \cdot p_b\right) }{ \left(p_a \cdot k_i\right) \left(p_b \cdot k_i\right)   }.
\end{align}
The factor $\CF$ in Eq. \eqref{eq:Def:WP-m} denotes the Casimir color operator in the fundamental representation; $\CF = (N_c^2-1)/2N_c$, with $N_c$ being the number of quarks.
The jet veto constraint $\Theta_i^\rho$ is given by: $\Theta_i^\rho = \Theta(\varrho_i - \rho)$. The simplicity of the eikonal amplitudes for primary emissions, as presented in Eq.~\eqref{eq:Def:WP-m}, facilitates the calculation of CLs to arbitrary loop orders in the perturbative series. However, the \kt clustering algorithm introduces non-trivial modifications to the final-state phase space, which complicates the computation.
The \kt jet algorithm, detailed extensively in Ref.~\cite{Cacciari:2011ma} and in our earlier works \cite{Delenda:2012mm, Khelifa-Kerfa:2024roc, Khelifa-Kerfa:2024hwx}, may be summarized in the following steps:
\begin{enumerate}
	\item Begin with a list of final-state particles, denoted as $I$.
	\item For each pair $(ij)$ in $I$, compute the distance measures:
	\begin{align}
		d_{ij} &= \min\left(k_{ti}^2, k_{tj}^2\right) \frac{\Delta R_{ij}^2}{R^2}, \qquad
		d_{iB} = k_{ti}^2,
	\end{align}
	where $\Delta R_{ij}^2 = \eta_{ij}^2 + \phi_{ij}^2$, with $\eta_{ij} = \eta_i - \eta_j$, and $\phi_{ij} = \phi_i - \phi_j$. $R$ is the jet-radius parameter and $B$ denotes the beam direction.
	\item If the smallest distance is $d_{ij}$, merge particles $(i, j)$ into a single pseudo-jet by summing their four-momenta using the E-scheme. If $d_{iB}$ is the smallest, declare particle $i$ as a final jet and remove it from $I$.
	\item Repeat steps 2 and 3 until $I$ is empty.
\end{enumerate}
Notably, the {\it clustering condition} for a particle pair $(i, j)$ is given by:
\begin{align}\label{eq:Def:ClusteringCond}
	\Delta R_{ij}^2 < R^2.
\end{align}
Thus, if two particles $i$ and $j$ are within a circle of radius $R$ in the rapidity--azimuth space, they will be clustered together by the \kt algorithm; otherwise, they will not.

In the subsequent sections, we compute the summation over $X$ in Eq.~\eqref{eq:Def:JetmassXsec-m} and integrate over the relevant phase space to derive the jet mass fraction.

\subsection{One and two loops}
\label{subsec:1-2Loops}

For the emission of a single soft gluon, $k_1$, all jet algorithms, including $k_t$, leave the phase space unaffected. Clustering effects only arise with the emission of at least two gluons. At this order, there are precisely two gluon configurations: $X = R$ (real gluon) and $X = V$ (virtual gluon). Using the measurement operator rules outlined in \cite{Khelifa-Kerfa:2024roc} and summarized above, we find:
\begin{align}\label{eq:1loop:uWX1}
 \sum_\X \uh_1\, \W_1^\X = \uh_1 \W_1^\R + \uh_1 \W_1^\V = -\Theta_1^\inn\, \Theta_1^\rho\,\W_1^\R,
\end{align}
where, at this loop order, $\Uh_1 = \uh_1 = 1- \Theta_1^\R \Theta_1^\inn \Theta_1^\rho$. The step function $\Theta_1^\inn = \Theta\left(R^2 - \Delta R_{1j}^2\right) = 1$ if gluon $k_1$ ends up  inside the measured jet, $j$,  after applying the jet algorithm and zero otherwise. The one loop {\it clustering} function is then given by:
\begin{align}\label{eq:1loop:ClusFunc}
 \Xi_1(k_1) = \Theta_1^\inn.
\end{align}
Substituting \eqref{eq:1loop:uWX1} back into the expression of the jet mass fraction \eqref{eq:Def:JetmassXsec-m} (for $m=1$) we find, at SL accuracy,
\begin{align}\label{eq:1loop:S1}
 \S_1(\rho) &= \int \d\Phi_1\, \Xi_1\,\Theta_1^\rho\,\W_1^\R, \notag\\
&= -\asb\,\CF \int_0^1 \frac{\d x_1}{x_1} \int_{-\infty}^{+\infty} \d\eta_1 \int_0^{2\pi} \frac{\d\phi_1}{2 \pi}\, w^1_{ab}\times \notag \\
& \times \Theta\left[2 x_1 \left(\cosh\eta_1 - \cos\phi_1 \right) - \rho\right] \Theta\left(R^2 - \eta_1^2 - \phi_1^2\right),  \notag\\
&= -\asb\,\CF\, \frac{L^2}{2},
\end{align}
where $L = \ln(R^2/\rho)$.
The exponentiation of $\S_1$ gives the well-known Sudakov form factor, which resums leading logarithms. Its expansion manifests at each order in the perturbative distribution of the jet mass fraction, corresponding to the case where {\it all} gluons are clustered into the measured jet. It is now well established that this form factor misses crucial contributions for non-global observables, starting at two loops. These missing contributions include mainly NGLs and CLs. In this paper, we focus exclusively on determining the missing CLs contributions for the single-jet mass observable up to six loops.

At two loops, the integrand in \eqref{eq:Def:JetmassXsec-m} (for $m=2$) reads \cite{Khelifa-Kerfa:2024roc}:
\begin{align}\label{eq:2loop:uWX2}
 \sum_\X \Uh_2 \W_{12}^\X = \Theta_2^\inn \left[-1 + \Theta_1^\out \, \Ob_{12}\right] \Theta_1^\rho \Theta_2^\rho \W_{12}^{\R\R}.
\end{align}
where $\Ob_{ij} = 1-\O_{ij} = \Theta\left(d_{ij} - d_{jB}\right)$. The above clustering function (the product of $\Theta_2^\inn$ and the square bracket) becomes:
\begin{align}\label{eq:2loop:ClusFun-Full}
 \Xi_2^{\ktt}(k_1, k_2) = -\Theta_1^\inn \, \Theta_2^\inn - \Theta_1^\out \Theta_2^\inn \, \O_{12}
 = - \prod_{i=1}^2 \Xi_i - \oXi_{12}^{\ktt},
\end{align}
where we have replaced $-1$ in the square bracket in \eqref{eq:2loop:uWX2} by $-\Theta_1^\inn - \Theta_1^\out$ and simplified the resultant expression to arrive at \eqref{eq:2loop:ClusFun-Full}.
The first term in the above equation is the product of the one loop clustering function \eqref{eq:1loop:ClusFunc}, while the second term represents the two loop (irreducible) clustering function:
\begin{align}\label{eq:2loop:ClusFun-irred}
	\oXi_{12}^{\ktt} = \Theta_1^\out \Theta_2^\inn \, \O_{12}.
\end{align}
Substituting back into the formula for the jet mass fraction, we find, at two loops:
\begin{align}\label{eq:2loop:S2}
	\S_2^{\ktt}(\rho) = \int_{x_1 > x_2} \d\Phi_{1} \d\Phi_2 \left[ \Xi_1 \Xi_2 + \oXi_{12}^{\ktt} \right] \Theta_1^\rho \Theta_2^\rho \W_{12}^{\R\R}.
\end{align}
In the first term above, one can relax the energy ordering, $x_1 > x_2$, and multiply by $1/2!$, resulting in the integral factorization:
\begin{align}
	\frac{1}{2!} \int \d\Phi_1 \Xi_1\, \Theta_1^\rho \W_1^\R \, \int \d\Phi_2 \Xi_2\, \Theta_2^\rho \W_2^\R = \frac{1}{2!} \left(\S_1\right)^2.
\end{align}
For the second term, the energy integrals factorize, yielding $L^2/2!$, and the CLs contribution can be written as:
\begin{align}\label{eq:2loop:S2-CLs}
	\oS_2^{\ktt}(\rho) = +\asb^2 \frac{L^2}{2!}\, \CFsq\, \F_2(R),
\end{align}
where the two loop CLs coefficient, $\F_2$, is given by:
\begin{align}\label{eq:2loop:F2-A}
	\F_2(R) &= \int_{1} \int_{2} \oXi_{12}^{\ktt}\, w_{ab}^1 w_{ab}^2
	\notag\\
	&= \int_{-\infty}^{+\infty} \d\eta_1 \d\eta_2 \int_0^{2\pi} \frac{\d\phi_1}{2\pi} \frac{\d\phi_2}{2\pi}\,
	\Theta\left(\eta_1^2+\phi_1^2 - R^2\right)
	\notag\\&\quad \times \Theta\left(R^2 - \eta_2^2 - \phi_2^2\right)
	\Theta\left(R^2 - \eta_{12}^2 - \phi_{12}^2 \right)
	\notag\\&\quad \times \frac{2}{\cosh^2\eta_1 - \cos^2\phi_1} \frac{2}{\cosh^2\eta_2 - \cos^2\phi_2},
\end{align}
where we have introduced, for brevity, the shorthand notations:
\begin{align}
	\int_i \equiv \int_{-\infty}^{+\infty} \d\eta_i \int_{0}^{2\pi} \frac{\d\phi_i}{2\pi}.
\end{align}
The integral above can be evaluated numerically using, for instance, the multidimensional \texttt{Cuba} library \cite{Hahn:2004fe}. The results are shown in Fig.~\ref{fig:F} (the blue dashed curve labeled "2loops"). The coefficient is positive and nearly constant over a wide range of $R$ (up to about 0.8), varying slowly for larger values. Notably, as $R$ approaches zero, the coefficient $\F_2(R)$ does not tend to zero as one might expect; instead, it saturates at a constant value of approximately 0.183. This reflects the {\it boundary} (or {\it edge}) effect seen in the calculations of NGLs (see, for example, Refs.~\cite{Dasgupta:2002bw, Khelifa-Kerfa:2011quw, Larkoski:2016zzc}). In both cases, i.e., CLs and NGLs, the integrals of their respective coefficients involve the boundary of the jet, which accounts for the relatively large constant factor at $R = 0$. This behavior persists across all orders in the perturbative expansion, as clearly illustrated in Fig.~\ref{fig:F}.

\subsection{Three loops}
\label{subsec:3Loops}

At three loop order the integrand in \eqref{eq:Def:JetmassXsec-m} (for $m=3$) is given by \cite{Khelifa-Kerfa:2024roc}:
\begin{align}\label{eq:3loop:uWX3}
\sum_\X \Uh_3 \W^\X_{123} &= \Theta_3^\inn \Big[1 - \Theta_1^\out \,\Ob_{13} -\Theta_2^\out \, \Ob _{23} + \notag\\
&\hspace{3em} + \Theta_1^\out \left(\Theta_2^\out + \Theta_2^\inn \O_{12} \right) \Ob_{13} \Ob_{23} \Big] \times \notag\\&\times \Theta_1^\rho \Theta_2^\rho  \Theta_3^\rho\W^{\R\R\R}_{123}.
\end{align}
Following the same procedure highlighted at two loops, specifically using the relations $\Theta_i^\inn + \Theta_i^\out = 1$ and $\O_{ij} + \Ob_{ij} = 1$ for $i, j = 1, 2$, one can straightforwardly show that the clustering term above can be expressed in a form analogous to that of two loops. That is:
\begin{align}\label{eq:3loop:ClusFun-Full}
\Xi_3^{\ktt}(k_1,k_2,k_3) &= \prod_{i=1}^3 \Xi_i + \sum_{\substack{ijk=1\\i<j}}^3 \oXi_{ij}^{\ktt}\, \Theta_k^\inn +  \oXi_{123}^{\ktt},
\end{align}
where the first two terms on the left-hand side of the above equation are reducible, i.e., they can be expressed in terms of the one and two loop clustering functions. The last term represents the new (irreducible) three loop clustering function, which is given by:
\begin{align}\label{eq:3loop:ClusFun-CLs}
\oXi_{123}^{\ktt}
&= \Theta_1^\out \Theta_3^\inn \left[ \Theta_2^\out \, \O_{13} \O_{23} + \Theta_2^\inn \O_{12} \left(-1 + \Ob_{13} \Ob_{23} \right)  \right].
\end{align}
The jet mass fraction \eqref{eq:Def:JetmassXsec-m} may then be shown to reduce to the form:
\begin{align}\label{eq:3loop:S3-Full}
\S_3^{\ktt}(\rho) = \frac{1}{3!} \left(\S_1\right)^3 + \S_1 \times \oS_{2}^{\ktt} + \oS_{3}^{\ktt},
\end{align}
where the CLs contribution at this order, $\oS_{3}^{\ktt}$, is given by:
\begin{align}\label{eq:3loop:S3-CLs}
\oS_{3}^{\ktt}(\rho) = -\asb^3\, \frac{L^3}{3!}\,\CFcub\, \F_3(R),
\end{align}
with the three loop CLs coefficient:
\begin{align}
\F_3(R) &= \int_{1} \int_2 \int_{3} \oXi_{123}^{\ktt}\, w_{ab}^1 w_{ab}^2 w_{ab}^3,
\notag\\
&= \Bigg[ \int_{1_\out} \int_{2_\out} \int_{3_\inn} \O_{13} \O_{23} +
\notag\\& \hspace{1em} + \int_{1_\out} \int_{2_\inn} \int_{3_\inn} \O_{12} \left( -1 +\Ob_{13} \Ob_{23}\right)
\Bigg] w_{ab}^1 w_{ab}^2 w_{ab}^3,
\end{align}
where we have additionally introduced the shorthand notation
\begin{align}
\int_{i_\inn} = \int_i \Theta_i^\inn, \qquad \int_{i_\out} = \int_i \Theta_i^\out.
\end{align}
The result of the integration as a function of the jet-radius $R$ is illustrated in Fig.~\ref{fig:F}. It is evident that $\F_3$ remains constant over almost the entire range of $R$ considered. For small values of $R$, we find $\F_3 \sim -0.052$, thus confirming our previous calculations \cite{Delenda:2012mm}. Notice that since $\F_3$ is negative, the CLs contribution at this order, Eq.~\eqref{eq:3loop:ClusFun-CLs}, is positive. Recalling that the NGLs contribution at this order is also positive \cite{Khelifa-Kerfa:2015mma, Khelifa-Kerfa:2024hwx}, we deduce that CLs and NGLs enhance each other, rather than cancel out as they did at two loops.
Moreover, the ratio $|\F_3/\F_2|$ ranges from approximately $28\%$ at $R \sim 0$ to about $18\%$ at $R = 1.2$. This highlights the dominance of the two loop contribution in the CLs  distribution. This observation will be further confirmed at higher loop orders, as we shall discuss in the next and later sections.

\subsection{Four loops}
\label{subsec:4Loops}

Similar to previous orders, we find for the integrand at four loops:
\begin{align}\label{eq:4loop:uWX4}
\sum_\X \Uh_4 \W^\X_{1234} &= \Theta_4^\inn \Big[
	-1 + \Theta_1^\out \,\Ob_{14} + \Theta_2^\out \, \Ob _{24}
\notag\\&
+ \Theta_3^\out \, \Ob _{34} - \Theta_1^\out \left(\Theta_2^\out + \Theta_2^\inn \O_{12} \right) \Ob_{14} \Ob_{24}
\notag\\&
- \Theta_1^\out \left(\Theta_3^\out + \Theta_3^\inn \O_{13} \right) \Ob_{14} \Ob_{34}
\notag\\&
- \Theta_2^\out \left(\Theta_3^\out + \Theta_3^\inn \O_{23} \right) \Ob_{24} \Ob_{34}
\notag\\&
+ \Theta_1^\out \left(\Theta_2^\out + \Theta_2^\inn \O_{12} \right) \times
\notag\\&
\times  \left(\Theta_3^\out + \Theta_3^\inn \left[\O_{23} + \O_{13} \Ob_{23} \right] \right) \times
 \notag\\&
 \times \Ob_{14} \Ob_{24} \Ob_{34} \Big] \Theta_1^\rho \cdots \Theta_4^\rho\, \W^{\R\R\R\R}_{1234}.
\end{align}
By replacing every occurrence of unity in the above expression with $\Theta_i^\inn + \Theta_i^\out$, so that we always have a product of step functions involving all four gluons, it is possible to extract the clustering function at this order. For instance, the first term, $-1$, in the square bracket is rewritten as
$
(\Theta_1^\inn + \Theta_1^\out) (\Theta_2^\inn + \Theta_2^\out) (\Theta_3^\inn + \Theta_3^\out) (\Theta_4^\inn + \Theta_4^\out),
$
which is then expanded. For the second term, $\Theta_1^\out$, we write
$
\Theta_1^\out (\Theta_2^\inn + \Theta_2^\out) (\Theta_3^\inn + \Theta_3^\out) (\Theta_4^\inn + \Theta_4^\out),
$
and similarly for the remaining terms. We find:

\begin{align}\label{eq:4lop:ClusFun-Full}
\Xi_4^{\ktt}(k_1,\dots,k_4) &= -\Bigg[
\prod_{i=1}^4 \Xi_i + \sum_{\substack{ijk\ell=1 \\i<j, k<\ell}}^4 \oXi_{ij}^{\ktt}\, \Theta_k^\inn \Theta_\ell^\inn +
\notag\\&
+ \sum_{\substack{ijk\ell=1 \\i<j<k}}^4 \oXi_{ijk}^{\ktt} \Theta_\ell^\inn + \sum_{\substack{ ijk\ell=1\\i<j,i<k<\ell}}^4 \oXi_{ij}^{\ktt} \oXi_{k\ell}^{\ktt} +
\notag\\
&+ \oXi_{1234}^{\ktt} \Bigg].
\end{align}
Similar to two  and three loop functions, the above four loop clustering function contains reducible parts and a single new irreducible CLs part. The latter reads:
\begin{align}\label{eq:4lop:ClusFun-CLs}
\oXi_{1234}^{\ktt} &= \Theta_4^\inn  \Theta_1^\out \Bigg[
\Theta_2^\out
\Big(\Theta_3^\out\, \O_{14} \O_{24} \O_{34} +
\Theta_3^\inn \Big[\O_{13} \O_{24} \times
\notag\\& \times \left(-1+\Ob_{14} \Ob_{34}\right) + \O_{23} \O_{14} \left(-1+\Ob_{24} \Ob_{34}\right) + \notag\\
&+ \O_{14} \O_{24} \left(-1+\Ob_{14} \Ob_{24} \Ob_{34}\right) \Big]  \Big)
+ \Theta_2^\inn\, \O_{12} \times
\notag\\
& \times \Big( \Theta_3^\out\, \O_{34} \left(-1+\Ob_{14} \Ob_{24}\right) +\Theta_3^\inn (-1+\Ob_{13} \Ob_{23}) \times  \notag\\& \times \left(-1+\Ob_{14} \Ob_{24} \Ob_{34}\right)  \Big)
\Bigg].
\end{align}
Substituting Eq. \eqref{eq:4loop:uWX4} back into \eqref{eq:Def:JetmassXsec-m}, performing the integrals and simplifying we find that the four loop jet mass fraction exhibits a similar pattern of ``an expansion of an exponential" observed at two and three loops. That is, it can be written as:

\begin{align}\label{eq:4loop:S4-Full}
\S_4^{\ktt}(\rho) &= \frac{1}{4!} \left(\S_1\right)^4 + \frac{1}{2!} \left(\S_1\right)^2 \times \oS_{2}^{\ktt} + \S_1 \times \oS_{3}^{\ktt} +
\notag\\
&+ \frac{1}{2!} \left(\oS_{2}^{\ktt}\right)^2 + \oS_{4}^{\ktt},
\end{align}
where the (new irreducible) four loop CLs contribution is of the form:
\begin{align}\label{eq:4loop:S4-CLs}
\oS_{4}^{\ktt}(\rho) = +\asb^4\, \frac{L^4}{4!}\,\CFfour\, \F_4(R),
\end{align}
with the four loop CLs coefficient given by:
\begin{align}\label{eq:4loop:F4}
\F_4(R) = \int_1 \int_2 \int_3 \int_4 \oXi_{1234}^{\ktt}\,  w_{ab}^1 w_{ab}^2 w_{ab}^3 w_{ab}^4.
\end{align}
It is plotted in Fig.~\ref{fig:F} as a function of $R$. The coefficient is positive and nearly constant over the entire range of $R$, with a value of approximately $0.022$. On average, the ratio $|\F_4/\F_3|$ is around $40\%$. The decremental pattern in the size of the CLs  coefficients at higher orders indicates the convergence of the CLs series.

It is worth noting that calculations of CLs coefficients in the literature stop at this order \cite{Delenda:2006nf, Delenda:2012mm, Khelifa-Kerfa:2024hwx}. In the following sections, we present state-of-the-art calculations of the CLs distribution, extending it, for the first time in the literature, to five and six loops.

\section{Five loops}
\label{sec:5loops}
The five loop integrand in Eq. \eqref{eq:Def:JetmassXsec-m} (for $m=5$) can be determined from the master formula in \cite{Khelifa-Kerfa:2024roc}. It reads:
\begin{align}\label{eq:5loop:uWX5}
\sum_\X \Uh_5 \W^\X_{1\dots 5} &= \Theta_5^\inn \Big[
	1 - \Theta_1^\out \,\Ob_{15} - \Theta_2^\out \, \Ob _{25} - \Theta_3^\out \, \Ob _{35}
	\notag\\&- \Theta_4^\out \, \Ob _{45} + \Theta_1^\out \left(\Theta_2^\out + \Theta_2^\inn \O_{12} \right) \Ob_{15} \Ob_{25}
	\notag\\
	&
	+ \Theta_1^\out \left(\Theta_3^\out + \Theta_3^\inn \O_{13} \right) \Ob_{15} \Ob_{35}
	\notag\\&
	+ \Theta_1^\out \left(\Theta_4^\out + \Theta_4^\inn \O_{14} \right) \Ob_{15} \Ob_{45}
	\notag\\
	&+ \Theta_2^\out \left(\Theta_3^\out + \Theta_3^\inn \O_{23} \right) \Ob_{25} \Ob_{35}
	\notag\\&
	+ \Theta_2^\out \left(\Theta_4^\out + \Theta_4^\inn \O_{24} \right) \Ob_{25} \Ob_{45}
	\notag\\&
	+ \Theta_3^\out \left(\Theta_4^\out + \Theta_4^\inn \O_{24} \right) \Ob_{35} \Ob_{45}
	\notag\\
	&- \Theta_1^\out \left(\Theta_2^\out + \Theta_2^\inn \O_{12} \right) \times
	\notag\\& \times  \left(\Theta_3^\out + \Theta_3^\inn \left[\O_{23} + \O_{13} \Ob_{23} \right] \right) \Ob_{15} \Ob_{25} \Ob_{35}
	\notag\\&
	- \Theta_1^\out \left(\Theta_2^\out + \Theta_2^\inn \O_{12} \right) \times
	\notag\\& \times \left(\Theta_4^\out + \Theta_4^\inn \left[\O_{24} + \O_{14} \Ob_{24} \right] \right) \Ob_{15} \Ob_{25} \Ob_{45}
	\notag\\&
	- \Theta_1^\out \left(\Theta_3^\out + \Theta_3^\inn \O_{13} \right) \times
	\notag\\& \times \left(\Theta_4^\out + \Theta_4^\inn \left[\O_{34} + \O_{14} \Ob_{34} \right] \right)
	\Ob_{15} \Ob_{35} \Ob_{45}
	\notag\\&
	- \Theta_2^\out \left(\Theta_3^\out + \Theta_3^\inn \O_{23} \right) \times
	\notag\\& \times \left(\Theta_4^\out + \Theta_4^\inn \left[\O_{34} + \O_{24} \Ob_{34} \right] \right) \Ob_{25} \Ob_{35} \Ob_{45}
	\notag\\&
	+ \Theta_1^\out \left(\Theta_2^\out + \Theta_2^\inn \O_{12} \right) \times
	\notag\\& \times \left(\Theta_3^\out + \Theta_3^\inn (\O_{23} + \Ob_{23} \O_{13} ) \right) \times
	\notag\\& \times \left(\Theta_4^\out + \Theta_4^\inn (\O_{34} + \Ob_{34} \O_{24} + \Ob_{34} \Ob_{24} \O_{14}  ) \right)
	\notag\\& \times \Ob_{15} \Ob_{25} \Ob_{45} \Ob_{45}
	\Big] \Theta_1^\rho \cdots \Theta_5^\rho\, \W^{\R\dots\R}_{1\dots 5}.
\end{align}
The clustering function may be deduced from the above expression, using analogous techniques to previous lower orders, to be:
\begin{align}\label{eq:5loop:ClusFun-Full}
\Xi_5^{\ktt}(k_1,\dots,k_5)
	&= \prod_{i=1}^5 \Xi_i
	+ \sum_{\substack{ijk\ell m=1 \\i<j,k<\ell<m}}^5 \oXi_{ij}^{\ktt}\, \Theta_k^\inn \Theta_\ell^\inn \Theta_m^\inn +
	\notag\\&
	+ \sum_{\substack{ijk\ell m=1\\i<j<k,\ell<m}}^5 \oXi_{ijk}^{\ktt} \Theta_\ell^\inn \Theta_m^\inn +
	\notag\\&
		+ \sum_{\substack{ijk\ell m=1 \\i<j,k<\ell}}^5 \oXi_{ij}^{\ktt} \oXi_{k\ell}^{\ktt} \Theta_m^\inn
	+ \sum_{\substack{ijk\ell m=1 \\i<j,\ell<m}}^5 \oXi_{ijk\ell}^{\ktt} \Theta_m^\inn +
	\notag\\&
	+\sum_{\substack{ijk\ell m=1 \\i<j,i<k<\ell}}^5 \oXi_{ij}^{\ktt} \oXi_{k\ell m}^{\ktt}
	+ \oXi_{12345}^{\ktt}.
\end{align}
The (irreducible) clustering function at five loops, $\oXi_{12345}^{\ktt}$, is cumbersome and will therefore be provided in an accompanying \texttt{Mathematica} notebook file, ``\texttt{Xi5-Xi6.nb}". Substituting Eq.~\eqref{eq:5loop:uWX5} back into the expression for the jet mass fraction yields a factorized form, similar to that at lower loop orders and resembling the structure of the clustering function above (Eq.~\eqref{eq:5loop:ClusFun-Full}). That is:
\begin{align}\label{eq:5loop:S5-Full}
\S_5^{\ktt}(\rho) &= \frac{1}{5!} \left(\S_1\right)^5 + \frac{1}{3!} \left(\S_1 \right)^3 \times \oS_{2}^{\ktt} + \frac{1}{2!} \left(\S_1 \right)^2 \times \oS_{3}^{\ktt} +
\notag\\
& + \S_1 \times \oS_{4}^{\ktt} + \oS_{2}^{\ktt} \times \oS_{3}^{\ktt} + \oS_{5}^{\ktt},
\end{align}
where the five loops CLs terms is written in the usual form:
\begin{align}\label{eq:5loop:S5-CLs}
 \oS_{5}^{\ktt}(\rho) &= -\asb^5 \frac{L^5}{5!}\,\CFfive\, \F_5(R),
\end{align}
with the  CLs coefficient at this order given by:
\begin{align}\label{eq:5loop:F5}
\F_5(R) &= \int_1 \int_2 \int_3 \int_4 \int_{5} \oXi^{\ktt}_{12345}\, \prod_{i=1}^5 w_{ab}^i.
\end{align}
The plot of $\F_5$ as a function of $R$ is shown in Fig.~\ref{fig:F}. It is observed that $\F_5$ is negative (thus making the overall CLs contribution, Eq.~\eqref{eq:5loop:S5-CLs}, positive) and nearly constant over the range of $R$ shown in Fig.~\ref{fig:F}, with a value of approximately $-0.013$. The ratio $|\F_5/\F_4|$ ranges from $58\%$ to $77\%$ across the values of $R$.

\section{Six loops}
\label{sec:6loops}

Similarly, the six loop integrand may be deduced from the formula in \cite{Khelifa-Kerfa:2024roc}. The resulting expression is lengthy and will be provided in the \texttt{Mathematica} file ``\texttt{Xi5-Xi6.nb}". The clustering function can then be computed in a straightforward manner, yielding an expression analogous to those of previous orders (Eqs.~\eqref{eq:2loop:ClusFun-Full}, \eqref{eq:3loop:ClusFun-Full}, \eqref{eq:4lop:ClusFun-Full}, and \eqref{eq:5loop:ClusFun-Full}). The full expression of the irreducible CLs clustering function, $\oXi_{1\dots 6}^{\ktt}$, is provided in the aforementioned \texttt{Mathematica} file.

\begin{figure}
	\centering
	\includegraphics[scale=0.55]{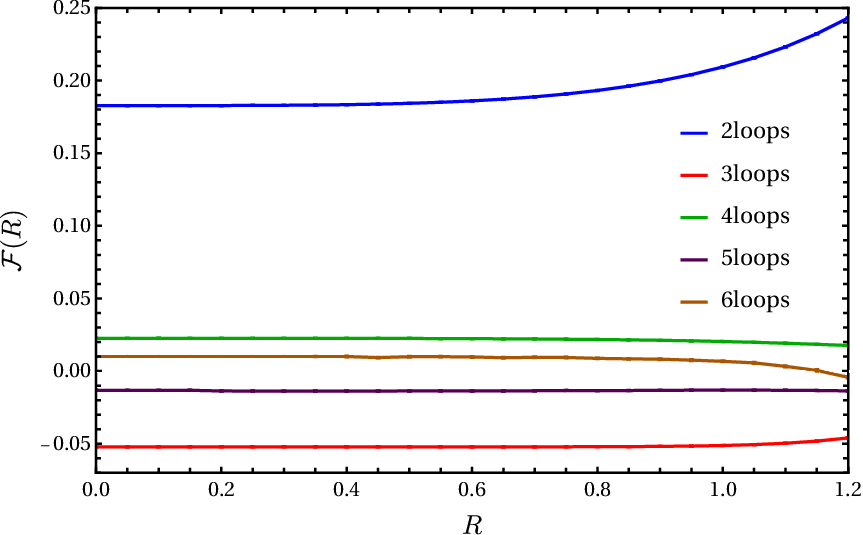}
	\caption{The CLs coefficients at two, three, four, five and six loops as a function of the jet-radius $R$.} \label{fig:F}
\end{figure}

The jet mass fraction may be shown to exhibit the usual factorized pattern:
\begin{align}\label{eq:6loop:S6-Full}
\S_6^{\ktt}(\rho) &= \frac{1}{6!} \left(\S_1 \right)^6 + \frac{1}{4!} \left(\S_1 \right)^4 \times \oS_{2}^{\ktt} + \frac{1}{3!} \left(\S_1 \right)^3 \times \oS_{3}^{\ktt} +
\notag\\
& +\frac{1}{2!} \left(\S_1\right)^2 \left[\frac{1}{2!} \left(\oS_{2}^{\ktt} \right)^2 +  \oS_{4}^{\ktt} \right] + \S_1 \times \oS_{5}^{\ktt} +
\notag \\
&+  \frac{1}{3!} \left(\oS_{2}^{\ktt}\right)^3 +  \frac{1}{2!} \left(\oS_{3}^{\ktt}\right)^2
+ \oS_{6}^{\ktt},
\end{align}
where the six loop CLs contribution is given in the usual form:
\begin{align}\label{eq:6loop:S6-CLs}
\oS_{6}^{\ktt}(\rho) = +\asb^6\, \frac{L^6}{6!}\,\CFsix\, \F_6(R),
\end{align}
with the CLs coefficient at this order given by:
\begin{align}\label{eq:6loop:F6}
\F_6(R) = \int_1 \int_2 \int_3 \int_4 \int_5 \int_{6} \oXi_{1 \dots 6}^{\ktt}\, \prod_{i=1}^6  w_{ab}^i.
\end{align}
 The above coefficient is plotted in Fig.~\ref{fig:F}. The six loop CLs coefficient, $\F_6(R)$, shares the same features as those of previous orders. In particular, it is nearly constant, positive with a value of approximately $0.010$. The ratio $|\F_6/\F_5|$ ranges from around $76\%$ to $30\%$ across the range of $R$.
%

\section{All-orders treatment}
\label{sec:AllOrders}

\subsection{Comparison to all-order numerical results}
\label{sec:MC-comparison}

The expressions for both the clustering functions and the jet mass fractions at each order up to six loops strongly suggest a pattern of exponentiation for the CLs distribution. However, at each higher order, there is a new CLs term that is not captured by the expansion of the exponential from previous orders. We may represent the said exponential using the following formula:
\begin{align}\label{eq:Allorders:Ct}
 \C(t) = \exp\left[ \sum_{n \geq 2} \frac{(-1)^n}{n!} \F_n(R)\, (2\,\CF\, t)^n \right],
\end{align}
such that, when expanded, $\C(t)$ reproduces the fixed-order distributions derived above. The evolution parameter, $t$, which governs the running of the coupling constant $\alpha_s$, is defined as:
\begin{align}\label{eq:Allorders:t}
 t = \frac{1}{2\pi} \int_{\frac{\rho}{R^2}}^1 \frac{\d x}{x} \, \as \left(x Q/2\right) = - \frac{1}{4 \pi \beta_0} \ln\left[1 - 2 \as \beta_0\, L\right],
\end{align}
where the second equality is the one loop expansion of $t$, with $\beta_0 = \left(11 \mathrm{C_A} - 2 n_f\right)/12 \pi$. Note that at fixed-order, $t = \bar{\alpha}_s L / 2$. Furthermore, although the signs of the CLs contributions in Eqs.~\eqref{eq:2loop:S2-CLs}, \eqref{eq:3loop:S3-CLs}, \eqref{eq:4loop:S4-CLs}, \eqref{eq:5loop:S5-CLs}, and \eqref{eq:6loop:S6-CLs} alternate, they are in fact all positive, as illustrated earlier.
To assess the convergence of the fixed-order series, its approximation to the all-orders distribution in Eq.~\eqref{eq:Allorders:Ct}, and the impact of the missing higher-order terms, we compare our results to the MC output of \cite{Dasgupta:2001sh}. Recall that this MC code resums single logarithms, initially for the anti-$k_t$ algorithm and later for the $k_t$ jet algorithm as well \cite{Appleby:2002ke, Delenda:2006nf}.
To extract the CLs form factor for direct comparison with Eq.~\eqref{eq:Allorders:Ct}, we run the MC for primary emissions only, with $k_t$ clustering turned ``on'' and ``off.'' The ratio of the two yields the CLs MC form factor, which we represent using the following parameterized form (analogous to the parametrization used for NGLs in the original paper \cite{Dasgupta:2001sh}):
\begin{align}
 \C_{\mathrm{\footnotesize MC}}(t) = \exp\left[ \CFsq\, \F_2(R)\left(\frac{1+(a t)^2}{1 + (b t)^c}\right) t^2 \right],
\end{align}
where the fitting parameters $a, b$ and $c$ are given by: $0.9, 3.12$ and $0.4$, respectively, for $R=0.7$. and $0.83, 3.16$ and $0.31$, respectively, for $R=1.0$.

\begin{figure}
	\centering
	\includegraphics[scale=0.55]{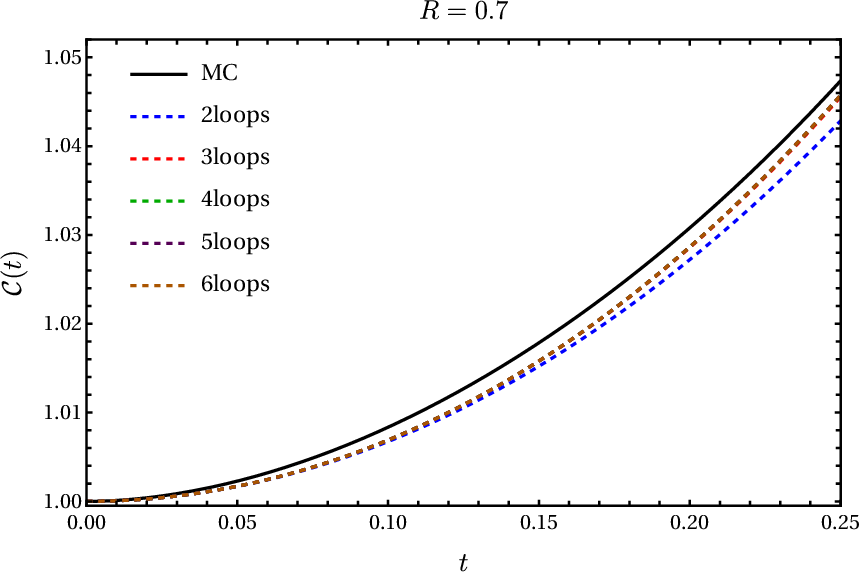} \vskip 1em
	\includegraphics[scale=0.55]{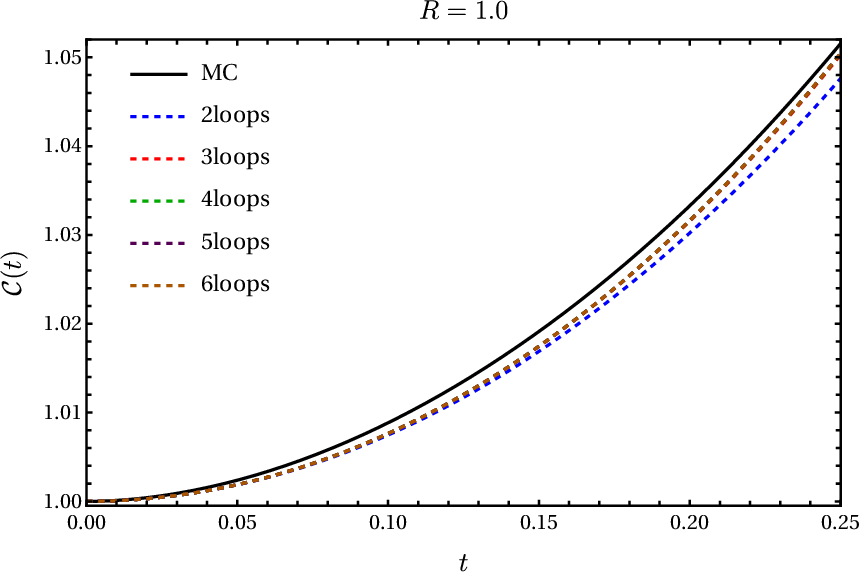}
	\caption{A comparison between the all-order MC CLs form factor and the analytical exponential \eqref{eq:Allorders:Ct}, as a function of the evolution parameter $t$.} \label{fig:Ct-exp}
\end{figure}

In Fig.~\ref{fig:Ct-exp}, we plot this form factor alongside the analytical exponential form in Eq.~\eqref{eq:Allorders:Ct} for two jet-radius values, namely $R = 0.7$ and $R = 1.0$. The labels ``2loops,'' ``3loops,'' and so forth indicate that the sum in the exponent of Eq.~\eqref{eq:Allorders:Ct} is truncated at $n=2$, $n=3$, and so on.

As observed in our previous work \cite{Delenda:2012mm}, the CLs form factor \eqref{eq:Allorders:Ct} is predominantly influenced by the two loop term, $\mathcal{F}_2$, which is both large (relative to higher-order terms) and suppressed only by a factor of $2!$. Higher-order terms, being smaller in magnitude and further suppressed by large factors of $n!$, have a minimal impact on $\mathcal{C}(t)$. In fact, curves for three loops and beyond coincide, indicating that differences among these terms in the exponential form are effectively unresolvable.
Note that the slight improvement observed for $R=1.0$ compared to $R=0.7$ is primarily driven by the larger value of $\mathcal{F}_2$, which increases slowly with the jet radius, as shown in Fig.~\ref{fig:F}. For the specific values of $t = 0.15$ and $t = 0.25$, the ratios of $\mathcal{C}(t)$, for $n=2$ and $n=6$, to the MC result are approximately $99.78\%$ and $99.84\%$, and $99.6\%$ and $99.9\%$, respectively, for $R=1.0$. This demonstrates that higher-order terms have a slightly more noticeable effect at larger values of $t$.

\begin{figure}
	\centering
	\includegraphics[scale=0.55]{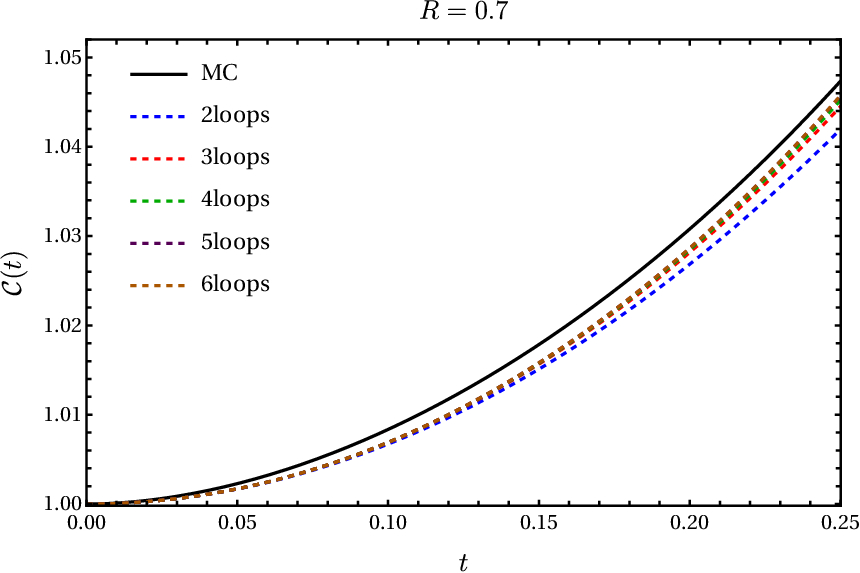} \vskip 1em
	\includegraphics[scale=0.55]{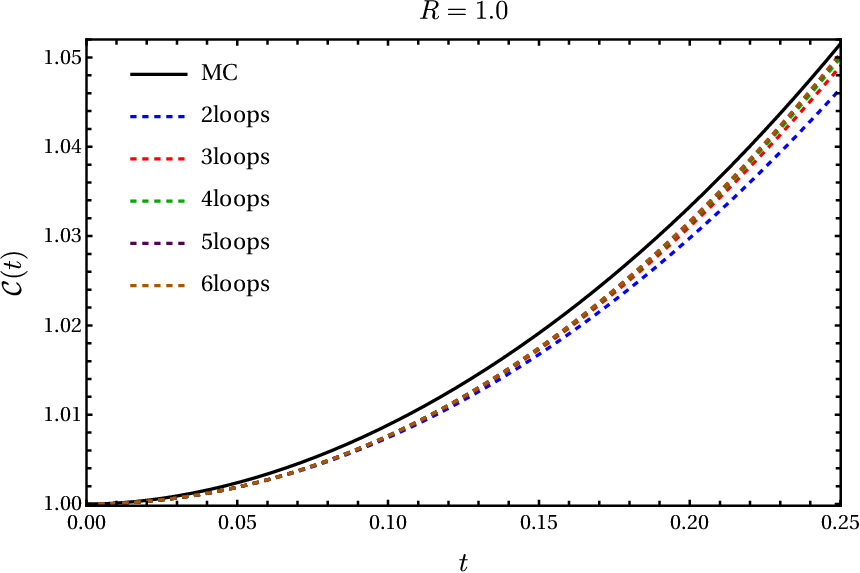}
	\caption{A comparison between the all-orders MC CLs form factor and the expansion of the analytical exponential \eqref{eq:Allorders:Ct-ser}, as a function of the evolution parameter $t$.} \label{fig:Ct-ser}
\end{figure}

In Fig. \ref{fig:Ct-ser} we compare the MC CLs form factor with the expansion of the exponential factor \eqref{eq:Allorders:Ct}, given by:
\begin{align}\label{eq:Allorders:Ct-ser}
 \C(t) &\simeq 1 + 2\,\F_2\,t^2 - \frac{4}{3}\,\F_3\,t^3 +  \frac{2}{3} \left(3 \F_2^2+\F_4\right) t^4 -
 \notag\\
 & - \frac{4}{15} \left(10 \F_2 \F_3 +\F_5\right) t^5 + \frac{4}{45} \Big(15 \F_2^3 + 10 \F_3^2 + \notag\\& + 15 \F_2 \F_4 + \F_6\Big) t^6.
\end{align}
The labels ``2loops", ``3loops" $\dots$ refer to truncations in the expression \eqref{eq:Allorders:Ct-ser} at $t^2, t^3$ and so on.
Unlike the exponential form \eqref{eq:Allorders:Ct}, the series representation of $\mathcal{C}(t)$ exhibits noticeable differences between the various loop contributions, particularly up to five loops and at larger values of $t$. The five and six loop curves are indistinguishable. For $R=1.0$ and $t = 0.15$ and $t = 0.25$, we find that the ratios of the series form \eqref{eq:Allorders:Ct-ser}, for $n=2$ and $n=6$, to the MC form factor are $99.76\%$, $99.84\%$, and $99.52\%$, $99.9\%$, respectively.

Overall, the analytical calculations, both the exponential and the series representations, provide excellent approximations to the all-orders numerical distribution for various jet radii and over a wide range of $t$ values. Given that the leading CLs contribution to the jet mass distribution is approximately $5\%$, we conclude that our results are sufficient for most phenomenological studies of interest.

\section{Conclusion}
\label{sec:Conclusion}

In this paper, we have presented, for the first time in the literature, the full structure of the leading clustering logarithms (CLs) up to six loops in perturbation theory for the $k_t$ jet algorithm. The formalism developed herein applies to a broad class of non-global logarithms arising in $e^+ e^-$ annihilation as well as in lepton-hadron and hadron-hadron collisions. As an example, we have explicitly demonstrated the calculations for the invariant single-jet mass observable.

Based on eikonal theory and imposing strong energy-ordering, the calculations are significantly simplified while retaining their validity up to single logarithmic accuracy. The master formula for the analytic structure of $k_t$ clustering, recently derived in Ref.~\cite{Khelifa-Kerfa:2024roc}, forms the backbone of the present work. The symmetric pattern of $k_t$ clustering enables the computation of higher-loop contributions in the perturbative expansion of generic non-global observables. While such calculations are feasible for primary abelian independent emissions, due to the simplicity of abelian eikonal amplitudes, they remain highly challenging for secondary non-abelian correlated emissions, given the inherent complexity of their corresponding eikonal amplitudes.

The computed CLs coefficients at two, three, four, five, and six loops share similar characteristics. Their dependence on the jet-radius parameter is weak (up to $R \approx 1.2$), and they approach a constant as $R \to 0$. This behavior has been previously associated with edge (or boundary) effects in the literature. Additionally, all coefficients are relatively small, with a suppression of $1/n!$ at each $n$-loop order. This smallness is reflected in the all-orders resummed form factor, which contributes at most $\mathcal{O}(5\%)$ of the full distribution (which includes both global logarithms and NGLs).

It was previously observed \cite{Delenda:2006nf, Delenda:2012mm} and further confirmed in this work that CLs exhibit a pattern of exponentiation. To evaluate the reliability and precision of the analytical exponential form and its series expansion, as well as the impact of missing higher-order coefficients, comparisons to all-orders numerical results were performed. Overall, the agreement is excellent across a wide range of logarithmic and jet-radius values, achieving percent-level accuracy. This indicates that the analytical calculations are phenomenologically robust and that the impact of higher-order terms is negligible.

There are numerous directions for future work building upon the current study. These include computing CLs coefficients beyond six loops, deriving an evolution equation for CLs analogous to that for NGLs, employing advanced techniques to probe the all-orders structure of CLs, considering next-to-leading CLs distributions, and generalizing the calculations to other jet algorithms, among others.


%

\newpage
\bibliography{Refs}

\end{document}